\documentclass[twoside]{dis07}
\usepackage[latin1]{inputenc}
\usepackage[dvips]{graphicx,epsfig,color}
\usepackage{wrapfig,rotating}
\usepackage{amssymb,amsmath,array}

\newcommand\beq{\begin{eqnarray}}
\newcommand\eeq{\end{eqnarray}}

\newcommand\la{\langle}
\newcommand\ra{\rangle}

\newcommand{\sslash}[1]{\not{\!#1}}

\def\GFt{\widetilde{G}_F}
\def\GDt{\widetilde{G}_D}
\def\gt{\widetilde{g}}

\begin{document}
%\title{DIS\,2007 Author Instructions}
\title{Factorization and Gauge Invariance of
Twist-3 Cross Section for Single Spin Asymmetry} 

%***********************************************************************
% AUTHORS INFORMATION AREA
%***********************************************************************
\author{Yuji Koike$^1$ and Kazuhiro Tanaka$^2$
%
% Optional short acknowledgment: remove next line if non-needed
%\thanks{This is an optional funding source acknowledgment.}
%
% DO NOT MODIFY THE FOLLOWING '\vspace' ARGUMENT
\vspace{.3cm}\\
%
% Addresses and institutions (remove "1- " in case of a single institution)
1- Department of Physics, Niigata University, 
Ikarashi, Niigata 950-2181, Japan
%
% Remove the next three lines in case of a single institution
\vspace{.1cm}\\
2- Department of Physics, Juntendo University, 
Inba-gun, Chiba 270-1695, Japan\\
}
%***********************************************************************
% END OF AUTHORS INFORMATION AREA
%***********************************************************************

\maketitle

\begin{abstract}
We prove factorization and gauge invariance of the twist-3
single-spin-dependent cross section 
in the leading order perturbative QCD,
which was missing in the previous literature.  
We emphasize that the consistency relation from the Ward identities for color gauge invariance
is crucial to guarantee the cancelation among the various
gauge-noninvariant contributions in the cross section.  
This relation also proves the absence of the ``derivative" terms in the cross section
corresponding to the hard-pole and soft-fermion-pole contributions.  
Applying the formalism to SIDIS, 
$ep^\uparrow\to e\pi X$, we have derived the complete cross section 
associated with the twist-3 distribution for the transversely polarized nucleon.
\end{abstract}

%\section{Submission}

\vspace{0.5cm}

In the framework of the collinear factorization for hard inclusive processes, single
spin asymmetry (SSA) in semi-inclusive reactions
is a twist-3 observable and can be described by using certain
quark-gluon correlation functions on the lightcone.  
In our recent paper\,\cite{EKT06n}, we have establised the 
twist-3 formalism for SSA, providing a proof for factorization property
and gauge invariance of the single spin-dependent cross section in the leading order
perturbative QCD,
which was missing in the previous literature\,\cite{ET82}-\cite{KQVY06}.  
We also applied the method to derive the formula for
SSA in semi-inclusive deep inelastic scattering (SIDIS), $ep^\uparrow\to e\pi X$.
Here we shall state what had to be clarified
for the twist-3 calculation in the previous literature, refering the readers to \cite{EKT06n}
for the details.

There are two independent twist-3 distribution functions
for the transversely polarized nuleon. 
They can be defined in terms of the gluon's field strength $F^{\alpha\beta}$ as
\begin{eqnarray}
& &\hspace{-0.5cm}M_F^\alpha (x_1,x_2)_{ij}\nonumber\\
& &=\int {d\lambda\over 2\pi}\int{d\mu\over 2\pi}
e^{i\lambda x_1}e^{i\mu(x_2-x_1)}
\langle p\ S_\perp |\bar{\psi}_j(0)[0,\mu n]
{gF^{\alpha\beta}(\mu n)n_\beta}[\mu n, \lambda n]
\psi_i(\lambda n)|p\ S_\perp \rangle\nonumber\\
& &=
{M_N\over 4} \left(\sslash{p}\right)_{ij} 
\epsilon^{\alpha pnS_\perp}
{G_F(x_1,x_2)}+ 
i{M_N\over 4} \left(\gamma_5 \!\! \sslash{p}\right)_{ij} 
S_\perp^\alpha
{\GFt(x_1,x_2)}+ \cdots,
\label{twist3F}
\end{eqnarray}
where $\psi$ is the quark field,
$p$ and $S_\perp$ are, respectively, momentum and spin vectors of the nucleon.  
$p$ can be regarded as
light-like ($p^2=0$) in the twist-3 accuracy, 
$n^\mu$ is the light-like vector 
($n^2=0$) with $p\cdot n=1$, 
and the spin vector satisfies $S_\perp^2 = -1$, $S_\perp \cdot p=S_\perp \cdot n=0$.  
$[\mu n,\lambda n]$ is
the gauge-link operator. 
The twist-3 distributions can also be defined in terms of the transverse component 
of the covariant dervative
$D_\perp^\alpha =\partial^\alpha_\perp -igA^\alpha_\perp$ as
\begin{eqnarray}
& &\int {d\lambda\over 2\pi}\int{d\mu\over 2\pi}
e^{i\lambda x_1}e^{i\mu(x_2-x_1)}
\langle p\ S_\perp |\bar{\psi}_j(0)[0,\mu n]
{D_\perp^{\alpha}(\mu n)}[\mu n, \lambda n]
\psi_i(\lambda n)|p\ S_\perp \rangle\nonumber\\
& &\qquad=
{M_N\over 4} \left(\sslash{p}\right)_{ij} 
\epsilon^{\alpha pnS_\perp}
{G_D(x_1,x_2)}+ 
i{M_N\over 4} \left(\gamma_5 \!\! \sslash{p}\right)_{ij} 
S_\perp^\alpha
{\GDt(x_1,x_2)}+ \cdots. 
\label{twist3D}
\end{eqnarray}
By introducing the nucleon mass $M_N$, the 
four functions $G_F$, $\GFt$, $G_D$ and $\GDt$ are defined as
dimensionless. 
The ``$F$-type" functions
$\{G_F, \GFt\}$ and the ``$D$-type" functions
$\{G_D, \GDt\}$ are not independent.  They are related as\,\cite{EKT06}
\beq
G_D(x_1,x_2)&=&P{1\over x_1-x_2}G_F(x_1,x_2),
\label{FDvector}\\
\GDt(x_1,x_2)&=&\delta(x_1-x_2)\gt(x_1)+
P{1\over x_1-x_2}\GFt(x_1,x_2),
\label{FDaxial}
\eeq
where $\widetilde{g}(x)$ is a function 
expressed in terms of 
$\{G_F, \GFt \}$ and the twist-2 quark helicity
distribution $\Delta q(x)$~\cite{EKT06} . 
The relations 
(\ref{FDvector}) and (\ref{FDaxial})
indicate that one cannot find the partonic hard cross sections
for the $F$-type and $D$-type functions independently.
We use the ``less singular'' $F$-type functions as a complete set for the twist-3 distributions.

\begin{figure}
\hspace*{0.7cm}
\includegraphics[height=.18\textheight,clip]{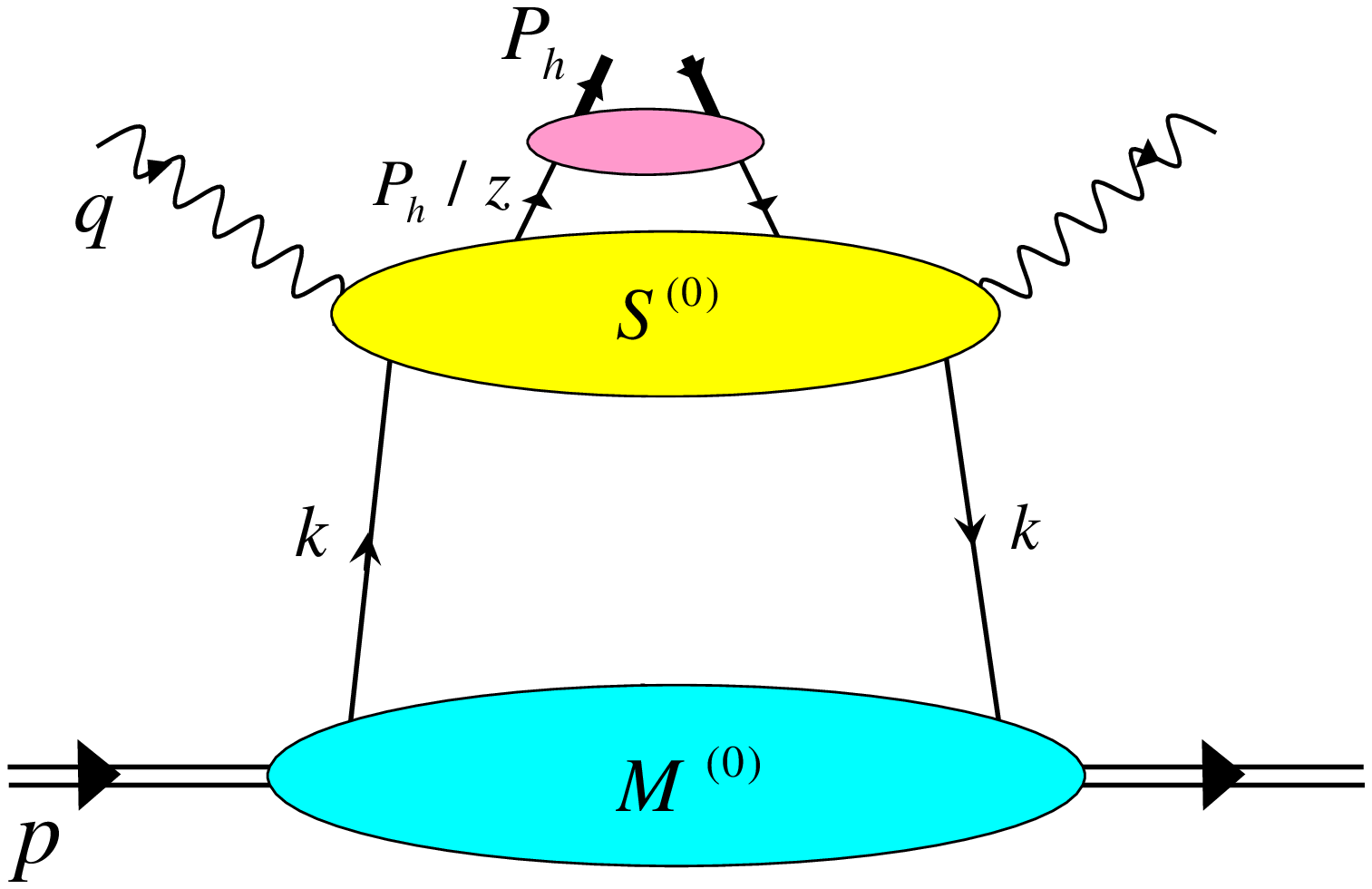}\hspace{1.6cm}
\includegraphics[height=.18\textheight,clip]{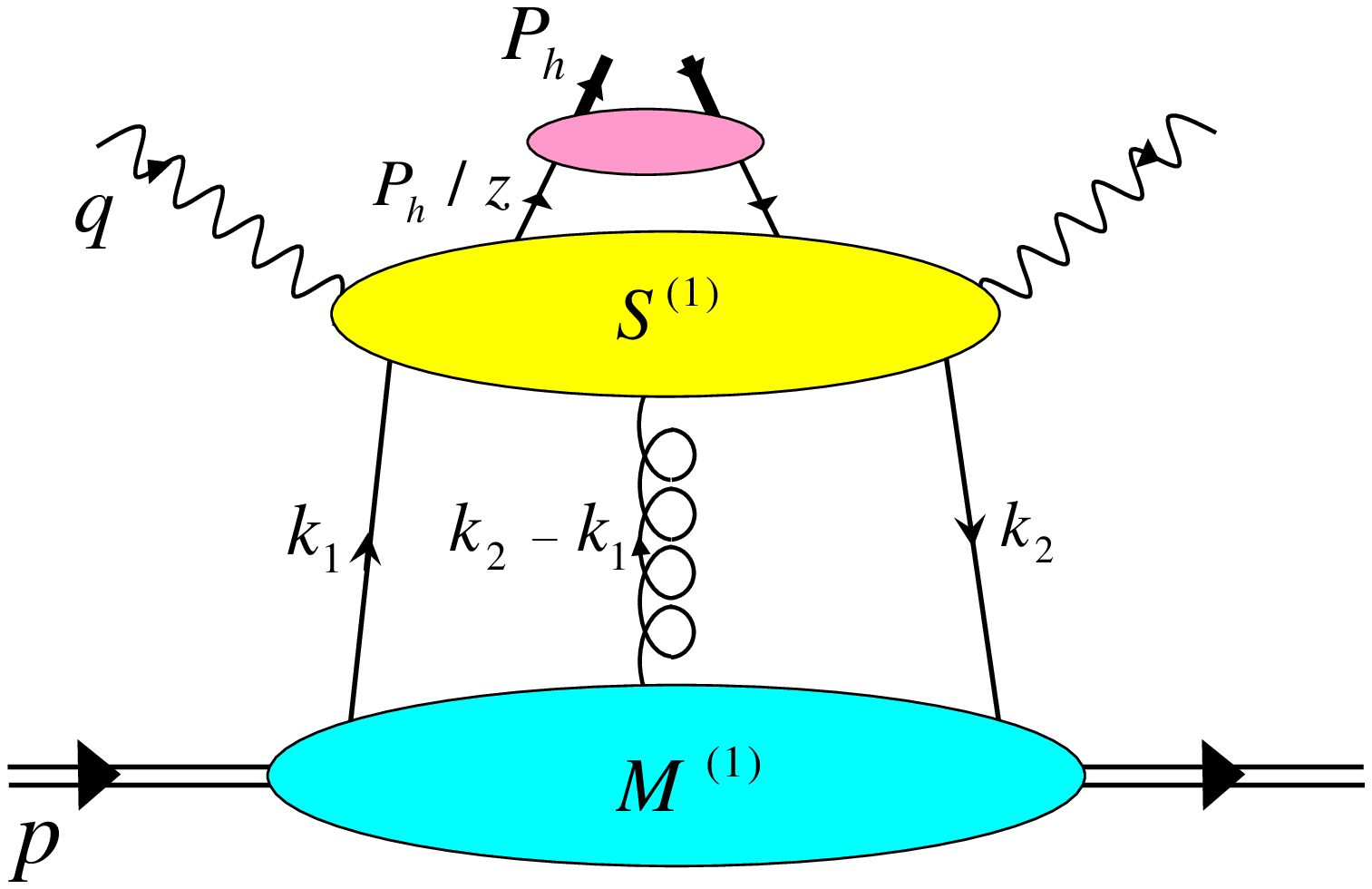}
\caption{Generic diagrams for the hadronic tensor of $ep^\uparrow\to e\pi X$
which contribute in the twist-3 accuracy.}
\end{figure}

To present our formalism we focus on the single-spin-dependent cross section
in SIDIS, $e(\ell)+p^\uparrow(p,S_\perp)\to e(\ell')+\pi(P_h)+X$, and consider the contribution
associated with 
the twist-3 distribution in the transversely polarized
nucleon.  Here the twist-2 fragmentation function for the final pion
is immediately factorized from the hadronic tensor 
$W_{\mu\nu}(p,q,P_h)$ ($q=\ell-\ell'$ is the momentum of the virtual photon) as
\beq
W_{\mu\nu}(p,q,P_h)=\sum_{j=q,g}\int{dz\over z^2}D_j(z) w^j_{\mu\nu}(p,q,{P_h\over z})\ ,
\label{Wmunu}
\eeq
where $D_j(z)$ is the quark and gluon fragmentation functions for the pion. 
To extract the twist-3 effect in $w_{\mu\nu}^j$, 
one has to analyze the diagrams
shown in Fig. 1, where the momenta for the parton lines are assigned.  
The lower blobs in the figure represent the nucleon matrix elements.  
They are the Fourier transforms of the correlation functions which are schmatically written as
$M^{(0)}(k)\sim \la \bar{\psi} \psi \ra$
and
$M^{(1)\alpha}(k_1,k_2)\sim \la\bar{\psi}A^\alpha \psi \ra$,
where $\la\cdots\ra\equiv \la p S_\perp|\cdots|pS_\perp\ra$.  
The middle blobs,
$S^{(0)}(k,q,P_h/z)$ and 
$S^{(1)}_\alpha(k_1,k_2,q,P_h/z)$, are the partonic hard scattering parts,
and the upper blobs represent the fragmentation function for the pion.  
To analyze these diagrams, 
we work in the Feynman
gauge.  
A standard
procedure to extract the twist-3 effect is the collinear expansion in terms of the
parton momenta $k$ and $k_{1,2}$ around the component pararell to
the parent nucleon's momentum $p$.
Since $M^{(1)\alpha=\perp}$
is suppressed by $1/p^+$ compared with 
$M^{(1)\alpha=+}$, one needs the collinear expansion 
only for the component $S^{(1)}_+= S^{(1)}_\rho p^\rho/p^+$ for the right diagram of Fig.~1.

In the approach by Qiu and Sterman\,\cite{QS91} using Feynman
gauge, 
$\partial^\perp A^+$ appearing in the collinear expansion is identified as
a part of the gluon's field strength $F^{\perp +}
=\partial^\perp A^+ -\partial^+ A^\perp -ig[A^\perp,A^+]$, and thus 
$\left.{\partial S^{(1)}_\rho(k_1,k_2,q,P_h/z)p^\rho
/ \partial k_{2\perp}^\alpha}\right|_{k_i=x_ip}$, appearing as the coefficient 
of $\la \bar{\psi}(\partial^\perp A^+)\psi\ra$,
was identified as 
the hard part for the $F$-type functions.  
In order to justify this procedure, it is necessary to
show that ``$-\partial^+A^\perp$" also appears with exactly the same coefficient as
that for $\partial^\perp A^+$. 
Note that, in the Feynman gauge, the matrix element of the type 
$\la  \bar{\psi}\partial^\perp A^+\psi  \ra$ 
is equally important as $\la   \bar{\psi}\partial^+A^\perp \psi \ra$, 
and both matrix elements have to be treated as the same order in the 
collinear expansion. This is because, even though
$\la  \bar{\psi}A^+\psi  \ra \gg \la  \bar{\psi} A^\perp \psi \ra$ 
in the Feynman
gauge 
in a frame with $p^+ \gg p^- , p^\perp$, 
action of $\partial^\perp$ to the gluon field in
$\la  \bar{\psi}A^+\psi  \ra $ brings relative suppression compared with 
$\partial^+$.  

On the other hand, $A^\perp$ was identified 
as a part of $D^\perp=\partial^\perp -ig A^\perp$ in \cite{QS91}, and  
$S_\perp^{(1)}(x_1p,x_2p,q,P_h/z)$, appearing as the coefficient 
of $\la \bar{\psi} A^\perp\psi\ra$, 
was treated as the hard part for the $D$-type functions,
independently from the $F$-type functions.  However, 
if one identifies
``$-\partial^+A^\perp$" part in $F^{\perp +}$, 
it also affects the coefficients of
$\la \bar{\psi} A^\perp\psi\ra$.  

In this way, the twist-3 formalism presented by \cite{QS91} was not complete,
in particular, the gauge invariance and uniqueness of factorization formula
was not shown explicitly.

Above consideration forces us to reorganize the collinear expansion.
Since $F$-type and $D$-type functions are related as in (\ref{FDvector}) and (\ref{FDaxial}),
we take the approach of expressing the cross section
in terms of $F$-type functions only.  
Detailed analysis of the diagrams shown in Fig. 1 shows that the single-spin-dependent
cross section in the leading order perturbation theory arises solely from the right diagram in
Fig.1.   
Taking into account the $-\partial^+A^\perp$ term in $F^{\perp +}$, we eventually arrive
at the following result for $w_{\mu\nu}^j$ in (\ref{Wmunu}):
\beq
w_{\mu\nu}^j& &
=\int\;dx_1\int\;dx_2{\rm Tr}\left[ i\omega^\alpha_{\ \ \beta}
M^{\beta}_{F} (x_1,x_2)
\left. {\partial S^{(1)}_\rho(k_1,k_2,q,P_h/z)p^\rho
\over \partial k_2^\alpha}\right|_{k_i=x_ip}
\right],
\label{final}
\eeq
with the consistency conditions
\beq
& &(x_2-x_1)\left.{\partial S^{(1)}_\rho(k_1,k_2,q,P_h/z)p^\rho
\over \partial k_2^\alpha}\right|_{k_i=x_ip}
+S_\alpha^{(1)}(x_1p,x_2p,q,P_h/z)=0,
\label{consistency1}\\
& &\left.{\partial S^{(1)}_\rho(k_1,k_2,q,P_h/z)p^\rho
\over \partial k_1^\alpha}\right|_{k_i=x_ip}
+\left.{\partial S^{(1)}_\rho(k_1,k_2,q,P_h/z)p^\rho
\over \partial k_2^\alpha}\right|_{k_i=x_ip}=0,
\label{consistency2}
\eeq
for $\alpha = \perp$, 
where $M^{\beta}_{F} (x_1,x_2)$ is given in (\ref{twist3F}), 
$\omega^{\alpha\beta}=g^{\alpha\beta}-p^\alpha n^\beta$ and  
${\rm Tr}[\cdots ]$ indicates the Dirac trace, while color trace is implicit.  
In (\ref{final}), the real quantity $w_{\mu\nu}^j$ is given as the
pole contributions\,\cite{ET82,QS91} from internal propagators of the hard part 
$\left. {\partial S^{(1)}_\rho(k_1,k_2,q,P_h/z)p^\rho
/ \partial k_2^\alpha}\right|_{k_i=x_ip}$, which are classified as
hard-pole (HP), soft-fermion-pole (SFP) and soft-gluon-pole (SGP) contributions.
Since these three kinds of poles occur at different points in phase space,
one can prove that each pole contribution satisfies the two conditions
(\ref{consistency1}) and (\ref{consistency2}) separately, 
and the whole hadronic tensor $w_{\mu\nu}^j$
can be written in terms of the $F$-type functions only,
including all the
HP, SFP and SGP contributions. 
We emphasize the crucial role of the two conditions (\ref{consistency1}) and (\ref{consistency2}):
%that, 
Using
these conditions 
%(\ref{consistency1}) and (\ref{consistency2}) 
%Because these are satisfied 
for each pole contribution, 
%the remaing terms in $w_{\mu\nu}^j$ 
the various terms associated with the gauge-noninvariant matrix elements,
arising in the collinear expansion of $w_{\mu\nu}^j$,
cancel out and only the term shown in (\ref{final}) remains, which
guarantees the factorization of the cross section and the gauge invariance of the formula. 
In particular, the first condition (\ref{consistency1}) can be proved using
the Ward identities for color gauge invariance (see \cite{EKT06n} for the detail). 

Pragmatically, the authors of \cite{QS91} reached the same formula (\ref{final}), but only
for the SGP contribution.  
We emphasize, however, that our argument leading to (\ref{final}) is totally different
from \cite{QS91}, in particular, (\ref{final}) is the principal formula
for
all HP, SFP and SGP contributions.  

For the HP and SFP contributions in which $x_1\neq x_2$, (\ref{consistency1}) can be rewritten as
\beq
\left.{\partial S^{(1)}_\rho(k_1,k_2,q,P_h/z)p^\rho
\over \partial k_2^\alpha}\right|_{k_i=x_ip}
={1\over x_1-x_2}S_\alpha^{(1)}(x_1p,x_2p,q,P_h/z).
\label{HP}
\eeq
Owing to this relation, one can obtain the relevant hard cross section in (\ref{final}) also through
$S_\perp^{(1)}(x_1p,x_2p,q,P_h/z)$, which is much simpler to calculate than the derivative
of the amplitude,
$\left. {\partial S^{(1)}_\rho(k_1,k_2,q,P_h/z)p^\rho
/ \partial k_2^\alpha}\right|_{k_i=x_ip}$.  
Since $S_\perp^{(1)}(x_1p,x_2p,q,P_h/z)$ does not contain any derivative,
the HP and SFP contributions do not appear as the derivative terms, such as those
$\propto d G_F(x_{bj},x)/dx$ or $d \GFt(x_{bj},x)/dx$ etc.
For the SGP contribution, one has to calculate the derivative $\left. {\partial S^{(1)}_\rho(k_1,k_2,q,P_h/z)p^\rho
/ \partial k_2^\alpha}\right|_{k_i=x_ip}$, because of the lack of the relation
(\ref{HP}) for $x_1 = x_2$, and it gives rise to the 
derivative terms with
$d G_F(x,x)/dx$.
For the HP and SFP contributions, one can also rewrite
the cross section in terms of $D$-type functions by using the relations
(\ref{FDvector}) and (\ref{FDaxial}) for $x_1 \neq x_2$, without any subtlety. 

We have applied the above formalism to SIDIS and derived~\cite{EKT06n} the
complete single-spin-dependent cross section for $ep^\uparrow\to e\pi X$ associated with
the twist-3 distributions for the transversely polarized nucleon,  
including all three kinds (HP, SFP and SGP) of pole contributions for
$G_F(x_1,x_2)$ and $\GFt(x_1,x_2)$.  

Our twist-3 formalism 
%method to show the factorization property and the gauge invariance
%of the single-spin-dependent cross section 
can be easily extended 
to other processes, such as Drell-Yan and direct photon productions, 
$p^\uparrow p\to\gamma^{(*)} X$, and the pion production
in $pp$ collision, $p^\uparrow p\to\pi X$,
which gives solid theoretical basis to the previously obtained
cross section formula\,\cite{QS91}-\cite{KQVY06}.

%\begin{theacknowledgments}
\vspace{0.3cm}
The work of K.T. was supported by the Grant-in-Aid for Scientific Research No. B-19340063.  
%\end{theacknowledgments}

\begin{footnotesize}

% ----------------------------------------------------------------------------

\end{footnotesize}

% ****************************************************************************
% END OF BIBLIOGRAPHY AREA
% ****************************************************************************

\end{document}